\documentclass[11pt,reqno,a4paper]{amsart}
\usepackage{amsmath,amssymb,dsfont,graphicx,cite}
\def\beq{\begin{equation}}
\def\eeq{\end{equation}}
\def\bea{\begin{eqnarray}}
\def\eea{\end{eqnarray}}

\newcounter{appendix}
\newcommand{\CP}{{\mathcal P}}
\def\ep{\epsilon}
\def\ri{{\rm{i}}}
\def\pa{\partial}
\begin{document}
\title[Multiscale reduction of discrete nonlinear Schr\"odinger equations]
{Multiscale reduction of discrete  \\ nonlinear Schr\"odinger equations} 
\author[D. LEVI, M. PETRERA and C. SCIMITERNA]
{D. LEVI${}^\diamond$, M. PETRERA${}^{\flat}$ and C. SCIMITERNA${}^{\flat}$}

\maketitle

{\footnotesize{

\centerline{\it ${}^\diamond$Dipartimento di Ingegneria Elettronica}
\centerline{\it Universit\`a degli Studi Roma Tre and Sezione INFN, Roma Tre}
\centerline{\it Via della Vasca Navale 84, 00146 Roma, Italy}
\centerline{e-mail: \texttt{levi@fis.uniroma3.it}}

\vspace{.3truecm}

\centerline{\it ${}^\flat$Dipartimento di Fisica}
\centerline{\it Universit\`a degli Studi Roma Tre}
\centerline{\it Via della Vasca Navale 84, 00146 Roma, Italy}
\centerline{e-mail: \texttt{petrera@fis.uniroma3.it}}
\centerline{e-mail: \texttt{scimiterna@fis.uniroma3.it}}

}}
\begin{abstract}

We use a discrete multiscale analysis to study the asymptotic integrability of differential-difference
equations. In particular, we show that multiscale perturbation techniques 
provide an analytic tool to derive necessary integrability conditions for two well-known discretizations of
the nonlinear Schr\"odinger equation.

\end{abstract}


\section{Introduction}
The  nonlinear Schr\"odinger (NLS) equation,
\beq
 \ri \partial_t f + \partial_{xx} f = \sigma |f|^2 f,\qquad f=f(x,t), \qquad
 \sigma =\pm 1 \,,\label{e0}
\eeq
 is a {\it universal}  
nonlinear integrable partial differential equation for models with weak
nonlinear effects.  Here $x$ is the spatial variable and $t$ is the time, while $\pa$ denotes differentiation with respect to its subscript.
It has been central for almost
fourty years in many different scientific areas  and 
it appears in several physical contexts, see for instance \cite{AS,APT,SS}. 

In  their pioneering work \cite{zs} Zakharov and Shabat proved its integrability
by solving its associated spectral problem. 
From the integrability of Eq. (\ref{e0}) it follows the existence of
infinitely many symmetries and conservation laws, and the solvability
of its associated  Cauchy problem. In correspondence with its symmetries one 
finds an infinite number of exact solutions, the solitons,
which, up to a phase, emerge unperturbed from
the interaction among themselves. 

On the other hand, also the problem of the discretization of the NLS equation has been the subject of an
intensive research. In literature, one may find
a few discretizations of the NLS equation. An integrable differential-difference equation
discretizing Eq. (\ref{e0}) has been found
by Ablowitz and Ladik \cite{al}. It reads
\bea
 \ri\partial_{t}f_{n}+\frac{f_{n+1}-2f_{n}+f_{n-1}}{2 h^2}  = \sigma
|f_{n}|^2 \frac{f_{n+1}+f_{n-1}}{2}\, , \label{nls1}
\eea
where $n \in \mathbb Z$
and $h$ is a real parameter
related to the space-discretization. Eq. (\ref{nls1}) admits a Lax pair and it has an infinite number of generalized symmetries and local conservation laws, which provide an infinite number of explicit  soliton solutions \cite{APT}.  As one easily see, in the limit
as $h \rightarrow 0$,  Eq. (\ref{nls1}) goes into the NLS equation (\ref{e0}).

From the point of view of the applications,  the most relevant differential-difference NLS equation is given by
\beq
\ri\partial_{t}f_{n}+\frac{f_{n+1}-2f_{n}+f_{n-1}}{2 h^2}
= \sigma |f_{n}|^2 f_{n}\, . \label{nls2}
\eeq

By introducing the parameter $s=0,1$, the discrete NLS equations (\ref{nls1}-\ref{nls2}) may be combined in the equation:
\beq
 \ri\partial_{t}f_{n}+\frac{(f_{n+1}-2f_{n}+f_{n-1}) \left ( 1- s \sigma h^2  |f_{n}|^2  \right)}{2 h^2} =\sigma |f_{n}|^2 f_{n}\, . \label{nls3}
\eeq
The case $s=1$ corresponds to Eq. (\ref{nls1}) while the case $s=0$ gives Eq. (\ref{nls2}).

Eq.  (\ref{nls2})
is one of the most studied lattice models (see for instance
\cite{APT,DN, Surv_Tsir,Surv_Flach,Eilbeck,Bose} and references therein). Its  study  has a long and
fascinating history, beginning in the 50's of the last century in solid state
physics with the Holstein's model for polaron motion
in molecular crystals \cite{holstein}. Later on it appears in
biophysics with the Davydov's model
for energy transport in biomolecules \cite{scott}. Among  the many recent
applications of Eq. (\ref{nls2}) let us just mention the theory of
Bose-Einstein condensates in optical lattices \cite{Bose} and semiconductors \cite{Surv_Tsir}.
Its 
continuous limit
goes again into the integrable NLS equation (\ref{e0}).
The discrete NLS equation (\ref{nls2}) possesses exact  discrete breathers 
solutions \cite{Surv_Flach} and just a few number of conserved quantities and symmetries are known. Numerical schemes have been used to
exhibit its chaotic behavior \cite{AOT}. A proof of its non-integrability, based on multiscale techniques, has been recently presented by the
authors in  \cite{epl}.

Multiscale analysis \cite{t1,t2}
is an
important perturbative method for finding approximate solutions to many  physical
problems by reducing a given partial differential equation
to a simpler equation, which can be integrable 
\cite{ce}. 
Multiscale expansions are structurally strong and 
can be applied to both integrable and non-integrable systems.
Zakharov and Kuznetsov \cite{zk} have shown that,
starting from an integrable partial differential equation and performing a proper multiscale
expansion, one may obtain other integrable systems. In particular, they showed
that the slow-varying amplitude of a dispersive wave solution of Eq. (\ref{e0}) satisfies the  Korteweg-de Vries (KdV) equation, and
viceversa.
 
Let us give a sketch of their derivation, showing how to obtain a KdV equation as the lowest order of the multiscale expansion of the NLS equation 
(\ref{e0}) with $\sigma=1$, the so called repulsive NLS equation. One separates the complex field $f$ in its amplitude and phase,
$f(x,t)=[\nu(x,t)/2]^{1/2} \exp 
[{\ri \phi(x,t)}]$, and rewrite the NLS equation as the system of real partial differential equations
\bea 
&&\pa_t \nu_t + \pa_x (\nu \varphi) = 0,  \label{kz1}\\ \label{kz2}
&&\pa_t \varphi + \varphi \pa_x \varphi   + \pa_x \nu  = \frac{1}{2} \pa_x \nu^{-1/2} \pa_x^2 \nu^{1/2},
\eea
where $\varphi= \pa_x \phi$. For long waves and small perturbations around the equilibrium solution we can define the following 
formal perturbation expansions:
\bea
\nu (x,t)&=& 1+ \sum_{i=1}^{\infty} \epsilon^{2i} \nu^{(i)} (x',t'),  \label{kz3} \\ \label{kz4}
\phi(x,t) &=& -t + \sum_{i=1}^{\infty} \epsilon^{2i-1} \phi^{(i)}(x',t'), 
\eea
where $x'= \epsilon (x-t)$ and $t'=\epsilon^3 t$ are suitable {\it slow}-variables, since $\ep$ is a small perturbation parameter.
By inserting expansions (\ref{kz3}-\ref{kz4}) into Eqs. (\ref{kz1}-\ref{kz2}), a direct computation shows that
the lowest nontrivial order of the perturbation, that is $\ep^3$, provides an evolution equation for the field $\nu^{(1)}$ w.r.t. the slow-time
$t'$:
\bea \nonumber
\pa_{t'} \nu^{(1)}+ \frac{3}{2} \nu^{(1)} \pa_{x'}\nu^{(1)}- \frac{1}{8} \pa_{x'}^3 \nu^{(1)}=0,
\eea
that is a KdV equation.

In \cite{ce} Calogero and Eckhaus have used the multiscale technique, at its lowest nontrivial order,
as a tool to give necessary conditions for the integrability of large classes of partial differential equations both in 1+1 and 2+1 dimensions. In particular it has been shown that the non-integrability of the resulting multiscale reduction is a consequence of the non-integrability of the ancestor system. 
The derivation of the higher order terms of multiscale expansions has been carried out by Degasperis, Manakov and Santini in 
\cite{DMS} and Kodama and Mikhailov in \cite{MK}.
In \cite{dp} Degasperis and Procesi introduced the notion of {\it{asymptotic integrability}}
of order $n$ by requiring that the multiscale expansion be verified up to a fixed order $n$ of the perturbation parameter. An integrable partial differential
equation, as the NLS equation (\ref{e0}), has an asymptotic integrability of infinite order.

Recently, some attempts to
extend this approach  to discrete equations have been
proposed \cite{Ag,lm,levi,lp,HLPS,HLPS2,HLPS3}. In
\cite{lm,levi,lp} a 
multiscale technique for dispersive
$\mathbb{Z}^2$-lattice equations has been developed.  
The main idea of the method used in \cite{lm,levi,lp}   is based on
dilation transformations of discrete shift operators. Up to our knowledge this has been carried out for the first time by
Jordan \cite{Jordan}.  
Let us illustrate the basic procedure in the case of a function $f_n: \mathbb{Z}\rightarrow \mathbb{C}$ depending only on one discrete index.
Let $T_n$ be the shift operator, $T_n f_n = f_{n+1}$, and $\Delta_n= T_n-1$ be the difference operator of order one.
The difference of order $j$ in a new discrete variable $n'$ is expressed in terms of an infinite
number of differences on the lattice of variable $n$ by means of the following
formula \cite{Jordan}:
\beq
\Delta^j_{n'}  f_{n'}= \sum_{i=0}^{j}  (-1)^{j-i}
{j \choose i} f_{n'+i} =\sum_{i=j}^\infty  {j!\over i!}  \sum_{k=j}^i
 \omega^k \mathfrak{S}_{i}^{k} \, \mathcal{S}_{k}^j \, \Delta^i_n  f_n,  \label{difInt}
\eeq
where $\omega$ is the ratio of the increment in the lattice of variable $n$ 
with respect to that of variable $n'$ and 
the coefficients $\mathfrak{S}_{i}^{k}$ and $\mathcal{S}_{k}^j$ are the Stirling numbers of the
first and second kind respectively. 

The Jordan formula (\ref{difInt}) implies that a rescaling of a lattice variable gives rise to nonlocal results. 
Therefore, to avoid the presence of infinite sums one needs to truncate the series (\ref{difInt}), namely to introduce
a {\it slow-varying condition}:
\bea \label{55}
\Delta_n^{p+1} f_n = 0\, ,
\eea
$p$ being a positive integer. A more general slow-varying condition has been recently introduced in  \cite{levitem}.

Indeed, in \cite{lm,levi,lp}, the multiscale analysis has been performed taking into account condition
(\ref{55}). As a consequence, the reduced discrete equations
turned out to be non-integrable even if the ancestor equation was integrable. 
However, as shown in \cite{HLPS,HLPS2,HLPS3}, 
if $p=\infty$ the reduced equations become  formally continuous and
their 
integrability may be properly preserved by the discrete multiscale procedure. 
In this way  multiscale techniques easily fit with
both difference-difference and differential-difference equations. The results contained in 
 \cite{HLPS,HLPS2,HLPS3} confirm a
discrete analog of the Zakharov-Kuznetsov claim \cite{zk}:
``if a nonlinear dispersive discrete equation is integrable
then its lowest order multiscale reduction is an integrable NLS
equation''. 

In this paper we present  the  multiscale 
perturbation analysis of Eqs. (\ref{nls1}-\ref{nls2}), thus extending to the discrete setting the approach 
used in \cite{zk,DMS,dp,MK}.
The derivation of the higher order terms in the perturbation 
expansion will enable us to provide an analytic evidence of  the
non-integrability of Eq. (\ref{nls2}). In fact, even if its lowest order  reduction
is an integrable KdV-type equation, the higher orders reductions
exhibit non-integrable behaviors (see also our recent letter \cite{epl} where no details were presented).
 On the contrary, the same calculations for the case  of Eq. (\ref{nls1}) will show that the Ablowitz-Ladik
discrete NLS equation satisfies all the integrability conditions up to the same order considered in the non-integrable case.
This is an indication of its asymptotic integrability of finite order
but not a proof of its integrability as for it we should go up to infinite order.

The paper is organized as follows. Section \ref{Sec2}
is devoted to the presentation of some technical details and basic formulas for the multiscale analysis of Eqs. (\ref{nls1}-\ref{nls2}).
The main results of the perturbation analysis will be given in Section \ref{Sec3}. In the concluding Section \ref{concl} we discuss 
further perspectives of this approach.

\section{Basic formulas for multiscale analysis of discrete NLS equations} \label{Sec2}

As for the continuous NLS equation, also for the discrete NLS equation (\ref{nls3}) we introduce amplitude and phase
of the function $f_n(t)$, namely $f_n(t)= [\nu_n(t)]^{1/2} {\rm{exp}} [\ri \phi_{n}(t) ]$.
Therefore the discrete NLS equation (\ref{nls3})
may be written as the following nonlinear system of real differential-difference equations ($s=1$ for (\ref{nls1}) and $s=0$ for (\ref{nls2})):
\bea
&& \!\!\! \!\!\!  \!\!\!  \!\!\!  \!\!\!  \!\!\!   \partial_{t}\nu_{n}= \left(s \sigma \nu_n -  \frac{1}{h^2}  \right)
\left[ \sqrt{\nu_{n}\nu_{n+ 1}}\sin(\phi_{n+1}-\phi_{n})+\sqrt{\nu_{n}\nu_{n- 1}} \sin (\phi_{n- 1}-\phi_{n})\right],\label{nh1}\\
&&\!\!\! \!\!\!  \!\!\!  \!\!\!    \!\!\!  \!\!\!   \partial_{t}\phi_{n}= - \frac{1} {h^2}+\frac{1}
{2} \left[  \frac{1}{h^2}+  (s-2) \sigma \nu_n \right]
 \left[\sqrt{\frac{\nu_{n+ 1}}{\nu_{n}}}\cos (\phi_{n+ 1}-\phi_{n})+\sqrt{\frac{\nu_{n- 1}}{\nu_{n}}}\cos(\phi_{n- 1}-\phi_{n})\right].   \label{nh2} 
\eea

By analogy with the continuous case, see Eqs. (\ref{kz3}-\ref{kz4}),  the real fields $\nu_n(t)$
and $\phi_n(t)$  are expanded around the constant solution $f_n(t)=\exp{(-\ri\sigma t)}$ in the following way:
\bea
\nu_{n}(t)&=&1+\sum_{i=1}^{\infty}
\ep^{2i} \,\nu^{(i)}(\kappa, \{t_{m}\}_{m \geq 1})\, ,\label{exp1}\\                          
\phi_{n}(t)&=&-\sigma t+\sum_{i=1}^{\infty}
\ep^{2i-1} \,\phi^{(i)}(\kappa, \{t_{m}\}_{m \geq 1})\, ,\label{exp2}
\eea
where $\ep$, with $0 < \ep \ll 1$, is the perturbation parameter.
The fields $\nu^{(i)}$ and 
$\phi^{(i)}$ in Eqs. (\ref{exp1}-\ref{exp2}) depend on the slow-space variable
$\kappa=  \ep \zeta n$, $\zeta \in \mathbb R$, and the slow-time variables $  t_m = \ep^{2m-1}t$, $m \geq 1$. The free parameter
$\zeta$  will be fixed later so as to obtain a suitable continuous limit. 

In general, given a function $u_n(t) = v(\kappa, \{t_{m}\}_{m \geq 1})$ we expand  $u_{n \pm
  1}(t)$ and $\partial_t u_n(t)$ in terms of the slow 
 variables $\kappa$ and $\{t_{m}\}_{m\geq 1}$ (see \cite{HLPS,HLPS2} for further details).
 Let $T_n$ be the shift operator defined by $T_n^{\pm} u_n = u_{n \pm 1}$.
 Then we have:
\beq
u_{n \pm 1} =   \left(T_{\kappa}^{\pm}\right)^{\epsilon \zeta} v(\kappa, \{t_{m}\}_{m \geq 1}) =
\sum_{i=0}^\infty \frac{(\pm \ep  \zeta \delta_\kappa)^i}{i!}  v(\kappa, \{t_{m}\}_{m \geq 1})\, ,  \label{pdio}
\eeq
with
\beq
\delta_{\kappa} =\sum_{i=1}^\infty
\frac{(-1)^{i-1}}{i}\Delta_{\kappa}^i, \qquad \Delta_{\kappa}^i= (T_{\kappa}-1)^i, \label{mnb}       
\eeq
and
\beq
\partial_t  u_{n}= \sum_{i=1}^{\infty} \epsilon^{2i-1}\partial_{t_i}
v(\kappa, \{t_{m}\}_{m \geq 1}). \label{pdioa}  
\eeq

If  $u_n$ is  
a slow-varying function of order $p$, see Eq. (\ref{55}),
we can truncate the infinite series in Eq. (\ref{mnb}).
 In such a case the $\delta_\kappa$-operators  reduce
to polynomials in the $\Delta_\kappa$-operators of order at most $p$. 
Hereafter we shall assume $p=\infty$ and the
$\delta_\kappa$-operators 
are formal differential operators.

Taking into account the expansions (\ref{exp1}-\ref{exp2}) and Eqs. (\ref{pdio}) and (\ref{pdioa}) we have the following formulas for the
shifts of the functions $\nu_n(t)$ and $\phi_n(t)$,
\bea
 \nu_{n\pm
1}&=& 1+\sum_{j=2}^{\infty} \ep^j \sum_{i=1}^{\left[j/2\right]} \frac{(\pm \zeta \delta_\kappa)^{j-2i}}{(j-2i)!}
\nu^{(i)}(\kappa, \{t_{m}\}_{m \geq 1})\, ,\label{f1}\\
\phi_{n\pm 1} &=&-\sigma
t + \sum_{j=1}^{\infty} \ep^j\sum_{i=1}^{\left[(j+1)/2\right]}
 \frac{(\pm \zeta \delta_\kappa)^{j-2i+1}}{(j-2i+1)!}\phi^{(i)}(\kappa, \{t_{m}\}_{m \geq 1})\,  ,\label{f2}
\eea
and their time derivatives,
\bea
\partial_{t}\nu_{n} &=&\sum_{j=2}^{\infty} \ep^{2j-1}\sum_{i=1}^{j-1}\partial_{t_{i}}\nu^{(j-i)}(\kappa, \{t_{m}\}_{m \geq 1})\, ,\label{Pitagora1}\\
\partial_{t}\phi_{n} &=&-\sigma+\sum_{j=1}^{\infty} \ep^{2j} \sum_{i=1}^{j}\partial_{t_{i}}\phi^{(j-i+1)}(\kappa, \{t_{m}\}_{m \geq 1})\, .\label{Pitagora2}   
\eea

\section{Main results} \label{Sec3}

The multiscale analysis of the system of real differential-difference equations (\ref{nh1}-\ref{nh2}) is carried out
by inserting the formal expansions (\ref{exp1}-\ref{exp2}) and 
(\ref{f1}-\ref{Pitagora2}) into Eqs. (\ref{nh1}-\ref{nh2}) and by requiring that the resulting equations be satified at all orders in $\ep$. 

The lowest non-trivial order corresponds to $\ep^2$. It  gives
\beq
\nu^{(1)}=-\sigma \partial_{t_{1}}\phi^{(1)}. \nonumber
\eeq

From now on all results will be presented only for the phase functions $\phi^{(i)}$, $i \geq 1$, since
any $\nu^{(i)}$, $i >1 $, is obtained from the even perturbation orders and 
may be expressed in terms of the $\phi^{(j)}$'s with $j \leq i$ and their derivatives.

At order $\ep^3$ we get
$$
\left(\partial_{t_{1}}^2-c^2\delta_{\kappa}^2\right)\phi^{(1)}=0, \qquad c=\pm \frac{\zeta (\sigma- s h^2)^{1/2}}{h}.
$$
As $c$ has to be real our multiscale analysis is performed only for
$\sigma=1$. Note that $0 < h <1$. Moreover we choose $\zeta=h$, so that $c=\pm (1- s h^2)^{1/2}$; note that $c$ remains finite as
$h\rightarrow 0$. Therefore the resulting equation at this order is satisfied by
$\phi^{(1)}=  \phi^{(1)} (x , \{t_m\}_{m \geq 2})$ with $x= \kappa - c t_1$. Here we are assuming that the solution
is asymptotically bounded.

At order  $\ep^5$,  the no-secular term condition implies  
$\left(\partial_{t_{1}}^2-c^2 \delta_{\kappa}^2\right)\phi^{(2)}=0$, so that $\phi^{(2)}=  \phi^{(2)} (x , \{t_m\}_{m \geq 2})$. 
At this same order, the evolution equation for 
$\phi^{(1)}$ w.r.t. the slow-time $t_2$ reads
\beq \label{Caesar} 
\partial_{t_{2}}\phi^{(1)}= K_2\left[  \phi^{(1)}\right]\,, \qquad 
K_2\left[  \phi^{(1)}\right] = \alpha_1
 \partial_{x}^3\phi^{(1)} + \alpha_2 \left(\partial_{x}\phi^{(1)}\right)^2, 
\eeq
with 
\beq
\alpha_1 = \frac{c}{24}[3-(3s+1)h^2], \qquad \alpha_2 = s h^2 - \frac34. \nonumber 
\eeq
 Eq. (\ref{Caesar}) is 
 a potential KdV equation and  $K_2\left[  \phi^{(1)} \right]$ denotes exactly the second flow of the integrable hierarchy associated
 with the potential KdV equation. A necessary condition for the integrability of 
the system (\ref{nh1}-\ref{nh2}) is that its
multiscale reductions provide the integrable evolution equations ($j \geq 3$):
\beq \label{eq:hpkdv}
\partial_{t_{j}}\phi^{(1)}=K_j[\phi^{(1)}]= \beta_j \int^x du \,   \mathcal L^{j-1}
\left[\partial_{u}^2 \phi^{(1)}\right]   ,
\eeq
where $\mathcal L$ is the recursive operator associated with the KdV hierarchy,
$$
 \mathcal L \left[f(x)\right] = \partial_{x}^2 f(x)
-\frac{\partial_{x} \phi^{(1)}}{\alpha_1} f(x) - \frac{\partial_{x}^2 \phi^{(1)}}{2\alpha_1}
\int^x du f(u)  ,
$$
and the $\beta_j$'s are real coefficients to be fixed later.

According to a general procedure for the multiscale analysis of partial differential equations \cite{dp,DMS,degas} we now
 assign a formal degree to the $x$-derivatives of the functions 
$\phi^{(j)}$, 
$$ {\rm{deg}} \left(  \partial_x^\ell \phi^{(j)}\right)= \ell+
2j-1, \qquad \ell \geq 0,
$$
and  define  $\CP_n$  as the vector space spanned by the products of all 
derivatives $\partial_x^\ell \phi^{(j)}$ with total degree $n$. We denote
by $\CP_n^{(r)} \subset \CP_n$ the subspace spanned by those products of derivatives
$\partial_x^\ell \phi^{(j)}$ with
$j \leq r$. 

After caring for secularities, the order $\ep^7$ yields 
$\phi^{(3)}= \phi^{(3)} (x , \{t_m\}_{m \geq 2})$  and
the following non-homogeneous  evolution equation for the field $\phi^{(2)}$ w.r.t. the slow-time $t_2$, depending 
on $\phi^{(1)}$ and its derivatives:
\bea
 \partial_{t_{2}}\phi^{(2)}- \alpha_1 \partial_{x}^3\phi^{(2)}- 2 \alpha_2 \partial_{x}\phi^{(1)}\partial_{x}\phi^{(2)}&=&
-\partial_{t_{3}}\phi^{(1)} + \alpha_3 \left(\partial_{x}^2\phi^{(1)}\right)^2 + \alpha_4 \left(\partial_{x}\phi^{(1)}\right)^3 \nonumber \\
&&  + \,  \alpha_5 \partial_{x}\phi^{(1)}\partial_{x}^3\phi^{(1)}
+ \alpha_6 \partial_{x}^5\phi^{(1)}, \label{Romulus}
\eea 
where

\begin{center}
\vspace{.5truecm}
\begin{tabular}[h]{ll}
$\alpha_3 ={\displaystyle{ \frac{h^2 [16 h^2 s - 5 (1+ 3 s)] +7 }{64}}}$,&$ \qquad \alpha_4 = {\displaystyle{ \frac{ch^2 (1+ 7 s)}{12}}} $, \\
& \\
$\alpha_5=  {\displaystyle{ \frac{h^2 [16 h^2 s - 3 (3+ s)] -3 }{48}}},$ &$ \qquad \alpha_6=-
{\displaystyle{ \frac{c[ h^4(15s+1) + 30 h^2 (s-1) -15]}{1920}}}$.
\end{tabular}
\end{center}
\vspace{.5truecm}

Substituting Eq. (\ref{eq:hpkdv}) with $j=3$ into Eq. (\ref{Romulus}) and
fixing  $\beta_3=-\alpha_6$ in order
to remove residual secularities, Eq. (\ref{Romulus}) reduces to the following 
 evolution equation for the field $\phi^{(2)}$ w.r.t. the slow-time $t_2$:
\beq
\partial_{t_{2}}\phi^{(2)}-K_{2}^{\prime}\left[\phi^{(1)}\right]\phi^{(2)}=f^{(t_2)}\,
, \label{sd}
\eeq
where $K_{j}^{\prime}\left[\phi^{(1)}\right]\psi$ is the Fr\'echet derivative of the flow $K_{j}\left[\phi^{(1)}\right]$ along the direction $\psi$, 
$$
K_{j}^{\prime}[\phi^{(1)}] \, \psi =\frac{d} {dr}K_{j}[\phi^{(1)}+ r \psi] \Big|_{r=0}. 
$$

In Eq. (\ref{sd}) the forcing term $f^{(t_2)}$ is a well-defined element of $\CP_6^{(1)}$, ${\rm{dim}}\, \CP_6^{(1)}=3$, namely
 a linear combination of three independent differential monomials (see Appendix, Eq. (\ref{Attis1})), with known coefficients which
are polynomial functions of $h$. 

Now the request for integrability of (\ref{nh1}-\ref{nh2}) implies the existence of the following  evolution equation
for the field $\phi^{(2)}$ w.r.t. the slow-time $t_3$:
\beq
\partial_{t_{3}}\phi^{(2)}-K_{3}^{\prime}\left(\phi^{(1)}\right)\phi^{(2)}=f^{(t_3)}\,
,\label{Pretestato} 
\eeq
where $f^{(t_3)} \in \CP_8^{(1)}$, ${\rm{dim}}\, \CP_8^{(1)}=6$ (see Appendix, Eq. (\ref{Attis2})), so that
the following
compatibility condition must hold:
\beq
\left\{\partial_{t_{3}}-K_{3}^{\prime}\left[\phi^{(1)}\right]\right\}f^{(t_2)}=
\left\{\partial_{t_{2}}-K_{2}^{\prime}\left[\phi^{(1)}\right]\right\}f^{(t_3)}.\label{Caetani} 
\eeq

Such a condition allows us
to express the coefficients of the polynomial $f^{(t_3)}$ in terms of those of $f^{(t_2)}$ and it does not impose any further constraint
on the coefficients of $f^{(t_2)}$ (see Appendix, Eq. (\ref{mm})).
As  in our case this condition is satisfied, we  conclude that the nonlinear system (\ref{nh1}-\ref{nh2})
has an asymptotic integrability of order seven irrespective of the value of $s$.

The next perturbation order, that is $\ep^9$, gives rise to a bifurcation between the non-integrable ($s=0$)
and the integrable case ($s=1$). For the sake of clarity we will study separately the two cases.

\begin{itemize}

\item {\it{The case}} $s=0$.
At order $\ep^9$, the resulting equations provide the evolution of the field $\phi^{(3)}$ w.r.t. the slow-time $t_{2}$. 
This is given by an integro-differential equation. To reduce it to a purely differential equation we
introduce the fields $\varphi^{(j)}=\partial_{x}\phi^{(j)}$. Taking care of  secularities and taking into
account that $\phi^{(1)}$ evolves w.r.t. the slow-time $t_4$ according to Eq. (\ref{eq:hpkdv}) with $j=4$,  we get
$\phi^{(4)}=  \phi^{(4)} (x , \{t_m\}_{m \geq 2})$ and
\beq
\partial_{t_{2}}\varphi^{(3)}-H_{2}^{\prime}\left[\varphi^{(1)}\right]\varphi^{(3)}=g^{(t_2)}, \label{Angerona}
\eeq
where $H_{j}^{\prime}\left[\varphi^{(1)}\right]\psi$ is 
the Fr\'echet derivative along $\psi$ of the $j$-th KdV flow 
$H_{j}\left[\varphi^{(1)}\right]=\partial_x K_j \left[\varphi^{(1)}\right]$.
Here $g^{(t_2)}$ is a known element of the space $\mathcal P_{9}^{(2)}$, ${\rm{dim}}\, \CP_9^{(2)}=14$ (see Appendix, Eq. (\ref{Sutri1})).
The evolution equation of $\varphi^{(3)}$ w.r.t. the slow-time $t_{3}$ takes the form
\beq
\partial_{t_{3}}\varphi^{(3)}-H_{3}^{\prime}\left[\varphi^{(1)}\right]\varphi^{(3)}=g^{(t_3)}, \label{Tarqnas}
\eeq
where  the coefficients of 
$g^{(t_3)} \in \mathcal P_{11}^{(2)}$, ${\rm{dim}}\, \CP_{11}^{(2)}=31$, are determined by requiring the
compatibility condition 
\beq \label{56}
\left\{\partial_{t_{3}}-H_{3}^{\prime}\left[\varphi^{(1)}\right]\right\}g^{(t_2)}=
\left\{\partial_{t_{2}}-H_{2}^{\prime}\left[\varphi^{(1)}\right]\right\}g^{(t_3)}\, .
\eeq
Eq. (\ref{56}) is a necessary condition for the integrability of the system (\ref{nh1}-\ref{nh2}) with $s=0$. 
In this case only nine out of the fourteen coefficients of $g^{(t_2)}$ are
independent. Thus we have five integrability conditions (see Appendix for further details).
It turns out that the obtained constraints on the polynomial $g^{(t_2)}$ are not
satisfied by the coefficients computed in Eq. (\ref{Angerona}). 
Therefore the system (\ref{nh1}-\ref{nh2}) with $s=0$, namely the discrete NLS equation
(\ref{nls2}), does not fulfil the necessary conditions assuring its integrability.

\vspace{.3truecm}

\item {\it{The case}} $s=1$.
In this case the resulting equations are purely differential and one can remain within the potential KdV hierarchy. Taking care of  secularities and taking into account that $\phi^{(1)}$ evolves w.r.t. the slow-time $t_4$ according to Eq. (\ref{eq:hpkdv}) with $j=4$,  we get
$\phi^{(4)}=  \phi^{(4)} (x , \{t_m\}_{m \geq 2})$ and
\beq
\partial_{t_{2}}\phi^{(3)}-K_{2}^{\prime}\left[\phi^{(1)}\right]\phi^{(3)}=h^{(t_2)}\,
,\label{Angerona1}
\eeq
where $h^{(t_2)}$ is a known element of the space $\mathcal P_{8}^{(2)}$, ${\rm{dim}}\, \CP_8^{(2)}=11$.
The evolution equation of $\phi^{(3)}$ w.r.t. the slow-time $t_{3}$ takes the form
\beq
\partial_{t_{3}}\phi^{(3)}-K_{3}^{\prime}\left[\varphi^{(1)}\right]\phi^{(3)}=h^{(t_3)}\,
,\label{Tarqnas1}
\eeq
where  the coefficients of 
$h^{(t_3)} \in \mathcal P_{10}^{(2)}$, ${\rm{dim}}\, \CP_{10}^{(2)}=24$, are determined by requiring the
compatibility condition 
\bea \label{561}
\left\{\partial_{t_{3}}-K_{3}^{\prime}\left[\phi^{(1)}\right]\right\}h^{(t_2)}=
\left\{\partial_{t_{2}}-K_{2}^{\prime}\left[\phi^{(1)}\right]\right\}h^{(t_3)}\, .
\eea

In such a case it turns out that all the constraints imposed  by (\ref{561}) on the eleven coefficients of the polynomial $h^{(t_2)}$ are satisfied by the coefficients computed in Eq. (\ref{Angerona1}) (see Appendix for further details).
This proves that 
the system (\ref{nh1}-\ref{nh2}) with $s=1$, namely the discrete NLS equation
(\ref{nls1}), has an asymptotic integrability of order nine. Actually, since the  discrete NLS equation
(\ref{nls1}) is known to be integrable, its asymptotic integrability should be of order infinite.

\end{itemize}

The above results may be summarized in the following Proposition.

\vspace{.2truecm}
{\bf Proposition.} {\it{The nonlinear differential-difference equation
$$
\ri\partial_{t}f_{n}+\frac{f_{n+1}-2f_{n}+f_{n-1}}{2 h^2} = |f_{n}|^2 f_{n}\, , 
$$
is non-integrable. In particular, its multiscale reduction, carried out by using the formal expansions (\ref{exp1}-\ref{exp2}) and 
(\ref{f1}-\ref{Pitagora2}), shows that it has  an asymptotic integrability of order seven. 

The differential-difference Ablowitz-Ladik equation
$$
 \ri\partial_{t}f_{n}+\frac{f_{n+1}-2f_{n}+f_{n-1}}{2 h^2}=  |f_{n}|^2 \frac{f_{n+1}+f_{n-1}}{2}\, ,
 $$
has  an asymptotic integrability of order nine. (Actually its asymptotic integrability should be of order infinite since it is known to be integrable.)}}

\section{Concluding remarks} \label{concl}

The present paper has been devoted to the derivation of higher order terms of the multiscale
perturbation of discrete NLS equations around the constant equilibrium solution. This enabled us to study the asymptotic integrability of Eqs.
(\ref{nls1}-\ref{nls2}), thus proving that the discrete NLS equation (\ref{nls2}) is non-integrable. Such a result has been already
established in \cite{epl}, but a detailed presentation of the integrability conditions appears for the first time in the present paper.
Moreover, we have also investigated the asymptotic integrability of the Ablowitz-Ladik discrete NLS equation (\ref{nls1}). 

We notice that the obtained results can be also used to  construct approximate solutions of the discrete NLS equations (\ref{nls1}-\ref{nls2}). They will
be expressed in terms of the solutions of the continuous equations belonging to the KdV and potential KdV hierarchies. More precisely, the
solutions of the lowest order term of the multiscale expansion of  (\ref{nls1}-\ref{nls2})
will be expressed in terms of a soliton solution of the potential KdV equation.

It is worthwhile to notice that the presented discrete multiscale technique fits with both differential-difference and difference-difference 
equations. Therefore it can be used  to
investigate the asymptotic integrability of a large class of discrete dynamical systems.
The method turns out to be a useful analytic tool whenever one has to deal with a discrete equation whose integrability is not established yet. 

As a future work, we plan to investigate the asymptotic integrability of the following class of differential-difference equations \cite{ba}:
$$
\partial_{t}f_{n}+\frac{f_{n+1}-2f_{n}+f_{n-1}}{2 h^2}=
 f_n + g(f_{n-1},f_n, f_{n+1}),
$$
where $g$ is a homogeneous polynomial of degree three. The importance of the 
above class of equations lies in the fact that it admits translationally invariant discrete kinks solutions.



\section*{Appendix: The integrability conditions for the potential KdV and KdV hierarchies}

This Appendix is devoted to the presentation of the
integrability conditions for the potential KdV and KdV hierarchies we used in our derivation. In particular, the following formulas have been
used to obtain the results contained in Section \ref{Sec3}. Thus we shall use the same notation.

\vspace{.3truecm}

{\bf{The potential KdV hierarchy.}} 
The integrable hierarchy of the potential KdV equation is given in Eq. (\ref{eq:hpkdv}). The quantities
$K_{j}[\phi^{(1)}]$ and their corresponding linearizations $K_{j}^{\prime}[\phi^{(1)}] \psi$, for $j=2$ and $j=3$,
 read (here $\partial= \partial_x$)
\begin{eqnarray}
&&K_{2}\left[\phi^{(1)}\right]= \alpha_1 \partial^3 \phi^{(1)}+\alpha_2 \left(\partial \phi^{(1)} \right)^2,\label{Musmeci1}\\
&&K_{3}\left[\phi^{(1)}\right]=\beta_{3}\left\{\partial^5 \phi^{(1)}+\frac{5\alpha_2} {3\alpha_1}
\left[\frac{2\alpha_2} {3\alpha_1}
\left( \partial\phi^{(1)} \right)^3+\left( \partial^2 \phi^{(1)} \right)^2+2 \partial \phi^{(1)} \partial^3 \phi^{(1)}\right]\right\},\label{Musmeci2}
\eea
and
\bea
&&K_{2}^{\prime}\left[\phi^{(1)}\right]\psi=\alpha_1 \partial^3 \psi+2\alpha_2 \partial \phi^{(1)} \partial \psi,\nonumber\\
&&K_{3}^{\prime}\left[\phi^{(1)}\right]\psi=\beta_{3}\left\{\partial^5 \psi+\frac{10\alpha_2} {3\alpha_1}
\left[\partial\phi^{(1)} \partial^3 \psi
+\partial^2 \phi^{(1)} \partial^2 \psi + \frac{\alpha_2} {\alpha_1}
\left( \partial^2 \phi^{(1)}\right)^2\partial \psi+ \partial^3 \phi^{(1)} \partial \psi\right]\right\}. \nonumber
\end{eqnarray}
where $\alpha_1,\alpha_2, \beta_3$ are real coefficients (in our case they are polynomial functions of the parameter $h$).

The non-homogeneous terms $f^{(t_{2})} \in \mathcal P_6^{(1)}$, $f^{(t_{3})}\in \mathcal P_8^{(1)}$, given
 in Eqs. (\ref{sd}) and (\ref{Pretestato}) 
respectively, are
\begin{eqnarray}
f^{(t_{2})}&=&a_1\left(\pa\phi^{(1)}\right)^3+a_2 \partial \phi^{(1)}\partial^3\phi^{(1)}+a_3 \left(\partial^2\phi^{(1)}\right)^2,\label{Attis1}\\
f^{(t_{3})}&=&b_{1}\partial\phi^{(1)}\left(\pa^2\phi^{(1)}\right)^2+b_{2}\pa\phi^{(1)}\pa^5\phi^{(1)}+
b_{3}\pa^2\phi^{(1)}\pa^4\phi^{(1)}\nonumber\\
&&+ \, b_{4}\left(\pa \phi^{(1)}\right)^4+b_{5}\left(\pa \phi^{(1)}\right)^2\pa^3\phi^{(1)}+b_{6}\left(\pa^3\phi^{(1)}\right)^2.\label{Attis2}
\end{eqnarray}

The combatibility condition (\ref{Caetani}) implies the 
following algebraic relations between the coefficients $a_1,a_2,a_3$ and $b_1,...,b_6$:
\beq \label{mm}
\left\{
\begin{array}{l}
9 \alpha_1^2 b_1= 5\beta_{3}\left[9a_{1}\alpha_1+2\left(a_{2}+3a_{3}\right)\alpha_2\right] , \vspace{.2truecm}\\
3 \alpha_1 b_2 = 5\beta_{3}a_{2},\vspace{.2truecm} \\
3 \alpha_1 b_3 =5\beta_{3}\left(a_{2}+2a_{3}\right), \vspace{.2truecm}\\
54 \alpha_1^3 b_4 = 5\beta_{3}\alpha_2\left(27a_{1}\alpha_1-a_{2}\alpha_2\right),\vspace{.2truecm} \\
9 \alpha_1^2 b_5 =5\beta_{3}\left(9a_{1}\alpha_1+5a_{2}\alpha_2\right),\vspace{.2truecm}\\
3 \alpha_1 b_6 = 5\beta_{3}\left(a_{2}+a_{3}\right).
\end{array}
\right.
\eeq
The system (\ref{mm}) allows us to express the $b_i$'s as functions of the $a_i$'s without requiring any
constraints on the latter ones. This means that the compatibility condition (\ref{Caetani}) is satisfied for any $a_1,a_2,a_3$ provided
that (\ref{mm}) is fulfilled.

The non-homogeneous terms $h^{(t_{2})} \in \mathcal P_8^{(2)}$, $h^{(t_{3})}\in \mathcal P_{10}^{(2)}$, defined
 in Eqs. (\ref{Angerona1}) and (\ref{Tarqnas1}) respectively, are quite long, since ${\rm{dim}} \,  \mathcal P_8^{(2)} = 11$ and ${\rm{dim}} \,  \mathcal P_{10}^{(2)} = 24$. In order to present
the integrability conditions imposed by the compatibility equation (\ref{561}) it is sufficient to write down only the expression of 
$h^{(t_{2})}$. It reads
\begin{eqnarray}
h^{(t_{2})}&=& c_{1}\left(\pa^3\phi^{(1)}\right)^2+c_{2}\pa^2\phi^{(1)}\pa^4\phi^{(1)}+c_{3} \pa \phi_{x}^{(1)}\pa^5\phi^{(1)}+
c_{4} \pa \phi^{(1)}\left(\pa^2 \phi^{(1)}\right)^2\nonumber\\
&&+\, c_{5}\left(\pa \phi^{(1)}\right)^2\pa^3 \phi^{(1)}+c_{6}\left(\pa \phi^{(1)}\right)^4+
c_{7} \pa \phi^{(1)}\pa^3 \phi^{(2)}+c_{8} \pa^2\phi^{(1)} \pa^2 \phi^{(2)}\nonumber\\
&& +\, c_{9} \pa^3\phi^{(1)} \pa \phi^{(2)}+c_{10}\left( \pa \phi^{(1)}\right)^2 \pa\phi^{(2)}+c_{11}\left( \pa\phi^{(2)}\right)^2.\ \ \ \ \ \ \ \ \ \label{Blandino}
\end{eqnarray}
The compatibility condition (\ref{561}) allows one to express the twenty-four coefficients of $h^{(t_{3})}$ in terms of those of $h^{(t_{2})}$
(these algebraic relations are easy to derive but rather cumbersome and we do not present them here) and the following three
algebraic constraints involving the coefficients $c_1,...,c_{11}$ and the coefficients $a_1,a_2,a_3, \alpha_1,\alpha_2$ previously defined.
The obtained necessary integrability conditions read:
\begin{eqnarray}
c_{6}&=&\frac{\left[27a_{1}\left(a_{2}+4a_{3}\right)\alpha_1-\left(37a_{2}^{2}+46a_{2}a_{3}+12a_{3}^{2}\right)\alpha_2\right]c_{11}} {108\alpha_1^{2}\alpha_2}\nonumber\\
&& +\, \frac{\left[\left(17a_{2}+18a_{3}\right)\alpha_2-27a_{1}\alpha_1\right]c_{8}} {108\alpha_1^{2} }+
\frac{\left[\left(3a_{2}-8a_{3}\right)\alpha_2-3a_{1}\alpha_1\right]c_{9}} {36\alpha_1^{2}}\nonumber\\
&& +\, \frac{\left(18c_{2}-24c_{1}-55c_{3}\right)\alpha_2^{2}} {54\alpha_1^{2}} +\frac{\left(13c_{5}-3c_{4}\right)\alpha_2} {18\alpha_1}, \nonumber\\
c_ {7}&=&\frac{a_{2}c_{11}} {\alpha_2},\nonumber \\
c_ {10}&=& \frac{3a_{1}c_{11}} {\alpha_2}.\nonumber
\end{eqnarray}

\vspace{.3truecm}

{\bf{The KdV hierarchy.}} For the integrable hierarchy of the KdV equation we have
\begin{eqnarray}
&&H_{2}\left[\varphi^{(1)}\right]= \alpha_1 \pa^3 \varphi^{(1)} + 2\alpha_2 \varphi^{(1)}  \pa \varphi^{(1)}  ,\nonumber\\
&&H_{3}\left[\varphi^{(1)}\right]= \beta_{3}\left\{\pa^5 \varphi^{(1)}+\frac{10\alpha_2}  {3\alpha_1}\left[\frac{\alpha_2} {\alpha_1}
\left( \varphi^{(1)} \right) ^2 \pa \varphi^{(1)}+2 \pa \varphi^{(1)} \pa^2 \varphi^{(1)}+\varphi^{(1)} \pa^3\right]\right\},\nonumber
\eea
and
\bea
&&H_{2}^{\prime}\left[\varphi^{(1)}\right]\rho=\alpha_1 
\pa^3 \varphi^{(2)}+2\alpha_2 \left(\rho \pa \varphi^{(1)}+\varphi^{(1)} \pa \rho \right),\nonumber\\
&&H_{3}^{\prime}\left[\varphi^{(1)}\right]\rho=\beta_{3}\left\{ \pa^5 \rho+\frac{10\alpha_2}  {3\alpha_1}\left[\varphi^{(1)}\pa^3\rho+2 \pa\varphi^{(1)} \pa^2\rho+
\left(2 \pa^2 \varphi^{(1)}+\frac{\alpha_2} {\alpha_1}\left(\varphi^{(1)}\right)^2\right) \pa \rho\right.\right.\nonumber\\
&&\qquad \qquad \qquad \qquad \left.\left.+\left(\frac{2\alpha_2} {\alpha_1}\varphi^{(1)} \pa \varphi^{(1)}+ \pa^3 \varphi^{(1)}\right)\rho\right]\right\}.
\nonumber
\end{eqnarray}

The above expressions are obtained by differentiating with respect to $x$ the corresponding expressions 
given in Eqs. (\ref{Musmeci1}-\ref{Musmeci2}) and setting $\varphi^{(1)}=\partial_{x}\phi^{(1)}$, $\rho=\partial_{x}\psi$.

The non-homogeneous terms $g^{(t_{2})} \in \mathcal P_9^{(2)}$, $g^{(t_{3})}\in \mathcal P_{11}^{(2)}$, defined  in Eqs. (\ref{Angerona}) and (\ref{Tarqnas}) respectively, are quite long, since ${\rm{dim}} \,  \mathcal P_9^{(2)} = 14$ and ${\rm{dim}} \,  \mathcal P_{11}^{(2)} = 31$. In order to present
the integrability conditions imposed by the compatibility equation (\ref{56}) it is sufficient to write down only the expression of 
$g^{(t_{2})}$. It reads
\begin{eqnarray}
g^{(t_{2})}&=&d_{1} \pa^2\varphi^{(1)} \pa^3\varphi^{(1)}+
d_{2} \pa\varphi^{(1)}\pa^4\varphi^{(1)}+
d_{3}\varphi^{(1)}\pa^5\varphi^{(1)}+
d_{4} \left(\pa \varphi^{(1)} \right)^3\nonumber\\
&&+\, d_{5}\varphi^{(1)}  \pa \varphi^{(1)}\pa^2 \varphi^{(1)}+
d_{6}\left(\varphi^{(1)}\right)^2\pa^3\varphi^{(1)}+
d_{7}\left( \varphi^{(1)}\right)^3\pa \varphi^{(1)}+
d_{8} \varphi^{(1)}  \pa^3 \varphi^{(2)}\nonumber\\
&&+\, d_{9}\pa \varphi^{(1)} \pa^2\varphi^{(2)}+ 
d_{10}\pa^2\varphi^{(1)}\pa \varphi^{(2)}+
d_{11} \varphi^{(2)} \pa^3 \varphi^{(1)}+
d_{12} \left(\varphi^{(1)} \right)^2\pa \varphi^{(2)}\nonumber\\
&&+\, d_{13} \varphi^{(1)}  \varphi^{(2)} \pa \varphi^{(1)}+
d_{14} \varphi^{(2)} \pa \varphi^{(2)}.\label{Sutri1}
\end{eqnarray}
The compatibility condition (\ref{56}) allows to express the thirty-one coefficients of $g^{(t_{3})}$ in terms of those of $g^{(t_{2})}$
(these algebraic relations are easy to derive but rather cumbersome and we do not present them here) and the following five
integrability conditions involving the coefficients $d_1,...,d_{14}$ and the coefficients $a_1,a_2,a_3, \alpha_1,\alpha_2$ previously defined:
\begin{eqnarray}
d_{7}&=&\frac{\left[9a_{1}\left(12a_{3}+5a_{2}\right)\alpha_1-\left(45a_{2}^2+88a_{2}a_{3}+12a_{3}^2\right)\alpha_2\right]d_{14}} {54\alpha_1^2\alpha_2}\nonumber\\
&&+\, \frac{\left[\left(3a_{2}-8a_{3}\right)\alpha_2-3a_{1}\alpha_1\right]d_{10}} {9\alpha_1^2}+\frac{2\left[\left(21a_{3}+4a_{2}\right)\alpha_2-9a_{1}\alpha_1\right]d_{9}} {27\alpha_1^2}\nonumber\\
&&+\, \frac{\left(9d_{5}+8d_{6}-24d_{4}\right)\alpha_2 } {9\alpha_1}-\frac{2\left(12d_{1}-30d_{2}+85d_{3}\right)\alpha_2^2} {27\alpha_1^2},\nonumber\\
d_{8}&=&\frac{a_{2}d_{14}} {2\alpha_2},\nonumber \\
d_{11}&=&d_{10}-d_{9}+\frac{a_{2}d_{14}} {2\alpha_2},\nonumber \\
d_{12}&=&\frac{3a_{1}d_{14}} {2\alpha_2},\nonumber \\
d_{13}&=&\frac{3a_{1}d_{14}} {\alpha_2}. \nonumber
\end{eqnarray}


\end{document}